\documentstyle[12pt,aaspp4,fleqn]{article}

\lefthead{Nissim Kanekar}
\righthead{A critique of scaling behaviour in ....}
\begin{document}
\renewcommand{\thefootnote}{\fnsymbol{footnote}}
\title {A critique of scaling behaviour in non-linear 
structure formation scenarios.}
\author{ Nissim Kanekar}
\affil{ National Centre for Radio Astrophysics, TIFR, Pune-411007, India}
\authoraddr{National Centre for Radio Astrophysics, TIFR, \\
Post Bag 3, University of Pune campus,\\
 Pune-411007, India} 
\authoremail{nissim@ncra.tifr.res.in}
\begin{abstract}
\noindent Moments of the BBGKY equations for spatial correlation functions of 
cosmological density perturbations are used to obtain a differential 
equation for the evolution of the dimensionless function, $h = -
({v/{\dot{a}x}})$, where $v$ is the mean relative pair velocity. 
The BBGKY equations are closed using a hierarchical scaling ansatz 
for the 3-point correlation function. Scale-invariant solutions 
derived earlier by Davis and Peebles are then used in the non-linear regime, along with 
the generalised stable clustering hypothesis ($h \rightarrow $ const.), 
to obtain an expression for the asymptotic value of $h$, in terms
of the power law index of clustering, $\gamma$,and the tangential and radial
velocity dispersions. The Davis-Peebles solution is found to require that tangential
dispersions are larger than radial ones, in the non-linear regime; this can be
understood on physical grounds. Finally, stability analysis of the solution
demonstrates that the allowed asymptotic values of $h$, consistent with 
the stable clustering hypothesis, lie in the range $0 \leq h \leq 1/2$. 
Thus, if the Davis-Peebles scale-invariant solution (and the hierarchical model 
for the 3-pt function) is correct, the standard stable 
clustering picture ($h \rightarrow 1$ as $\overline\xi \rightarrow \infty$) 
is not allowed in the non-linear regime of structure formation. \\

\end{abstract}
\section{Introduction}
\noindent The cosmological BBGKY equations have been used on a number of occasions 
to study the evolution of non-linear density fluctuations in an expanding, flat 
($\Omega = 1$), background Universe (\cite{DP77}; \cite{RF}; \cite{YG}). These 
equations have the advantage of dealing directly 
with statistical quantities, {\it i.e.} the spatial N-point correlation functions; however, 
they form an infinite hierarchy in which the equation for the N-point correlation 
function contains terms involving the (N+1)-point function. This necessitates the 
development of closure schemes in which the hierarchy is cut off at a finite number of 
equations by making some assumptions wherein the higher order correlation functions 
are written in terms of the lower order ones. Such a scheme was used by Davis $\&$ 
Peebles (1977; hereafter DP) to demonstrate the existence of a similarity solution to 
the equations; the assumptions used were the stable clustering hypothesis, the 
vanishing of velocity skewness, a hierarchical model for the 3-point correlation 
function, $\zeta \propto <\xi^2>$, where $\xi$ is the 2-point correlation function, 
and finally, a factorisation of the 2- and 3-body phase space distributions 
which gives a form for the 3-body weighted pair velocity. In this solution, an 
initial power spectrum of form $P_o(k) \sim k^n$ evolves to give a 2-point 
correlation function\\
\setlength{\mathindent}{5.5 cm}
\begin{equation}
\xi(r) \propto r^{-\gamma}\mbox{,} \hskip 0.2 in \gamma = {{3(n+3)}\over {(n+5)}}
\end{equation}
\noindent in the strong clustering regime with $\xi \gg 1$. Ruamsuwan $\&$ Fry 
(1992; hereafter RF) showed that the assumption of vanishing velocity skewness was not a 
priori necessary but could be derived as a result of other, more general assumptions; 
the solution was also shown to be marginally stable to perturbations. Finally, 
Yano $\&$ Gouda (1997) found that the stability condition used by DP was satisfied 
for the unique case of vanishing skewness and the specific form for $\zeta$ in 
terms of the products of the 2-point correlation functions. The power index of 
the 2-point function in the strong clustering regime depends, in general, on 
the mean relative physical velocity, the skewness and the 3-point correlation function.
\par 
The present work considers the evolution of the function, $h 
\equiv -({{<v>}/ {\dot{a}x}})$, {\it i.e.} the ratio of the mean relative peculiar 
velocity to the Hubble velocity. The BBGKY equations are not derived again; instead,  
we use the relevant moment equations from RF to proceed. The form for the 3-point 
correlation function is the same as in DP and RF. We derive an equation for the 
evolution of $h$ in terms of the mean 2-point correlation function, $\overline\xi$, 
using the ansatz that $h$ is a function of $\overline\xi$ alone (\cite{Ham}; \cite{NP}, 
hereafter NP). This ansatz is, however, not crucial to the later 
discussion. (Note, further, that the above equation is derived using only 
the zeroth and first moments of the second BBGKY equation and hence does 
not contain any assumptions regarding the form of the velocity skewness 
or the higher velocity moments, excepting the fact that they be such as to yield the 
DP similarity solution for the two-point correlation function and the velocity 
dispersions; the assumption of vanishing velocity skewness is {\it not}
used here. Of course, as shown by RF, this assumption is not necessary to arrive 
at the DP solution; in fact, zero skewness results as only one particular case in 
their closure scheme.) The DP solution is then substituted in the equation for $h$ and the 
generalised stable clustering hypothesis ($h \rightarrow$ const., for $\overline\xi 
\gg 1$) used, to obtain an expression relating the asymptotic value of $h$ to $\gamma$, 
the power law index of clustering, and the radial and tangential pair velocity 
dispersions. This, and the requirement that $h$ is real, gives the constraint that 
tangential dispersions are larger than radial dispersions, in the non-linear regime. 
Finally, a stability analysis carried out by perturbing about the above solution 
shows that stable solutions are attained only for asymptotic values of $h$ in the 
range $0 < h < 0.5 $. Thus, the standard stable clustering picture ($h \rightarrow 
1$ as $\overline\xi \rightarrow \infty$) is incompatible with the DP scaling 
solution. \\
\par
\section{The BBGKY hierarchy}
\noindent The zeroth and first moments of the 2$^{\rm nd}$ BBGKY equation are 
respectively (RF, equations (22) and (23))\\
\setlength{\mathindent}{4.0 cm}
\begin{equation}
\label{initcontinuity}
{\partial\xi_{12}\over{\partial{t}}} + {1\over a}{\partial\over{\partial{x^i_{12}}}}
{\Big[}<v^i_{12}>(1+\xi_{12}){\Big]}= 0 
\end{equation}
\setlength{\mathindent}{0.0 cm}
\begin{displaymath}
{1\over a}{\partial\over{\partial{t}}}{\Big [}a<v^i_{12}>(1+\xi_{12}){\Big ]} 
+ {1\over a}{\partial\over{\partial{x^j_{12}}}}{\Big [}<v^i_{12}v^j_{12}>
(1+\xi_{12}){\Big ]}
\end{displaymath}
\setlength{\mathindent}{4.5 cm}
\begin{equation}
\label{bbgky2}
 + G{\rho_b}a\int d^3x_3(\xi_{13} + \xi_{23} + \zeta_{123}){\Bigg (} 
{{x^i_{13}}\over {\mid x_{13}\mid^3}} - {{x^i_{23}}\over {\mid x_{23}\mid^3}} 
{\Bigg )} = 0
\end{equation}
\noindent where $x^i_{12} \equiv x^i_1-x^i_2$ denotes the separation between 
particles labelled 1 and 2 and $v^i_{12} \equiv v^i_1-v^i_2$ is their relative 
velocity ($i$ is a vector index). $\xi_{12}$ and $\zeta_{123}$ are the usual 
2- and 3-point correlation functions respectively while $\rho_b$ is the 
background density.\\
\noindent The number of independent vector components in equation 
(\ref{bbgky2}) can be reduced by making use of various cosmological 
symmetries. The isotropy of the background Universe implies that 
the mean relative pair velocity, $<v^i_{12}>$, should be aligned with 
the pair separation, {\it i.e.} \\
\setlength{\mathindent}{6.0 cm}
\begin{equation}
<v^i_{12}> = v(x_{12})\hat{x}_{12}^i
\end{equation}
\noindent where $\hat{x}^i$ denotes a unit vector along $x^i$, and the pair velocity 
dispersion should have longitudinal and transverse polarisations about the mean
$\Pi$ and $\Sigma$ (RF).\\
\setlength{\mathindent}{4.5 cm}
\begin{equation}
<\Delta{v^i}\Delta{v^j}> = \hat{x}^i\hat{x}^j\Pi(x) + (\delta^{ij} - 
\hat{x}^i\hat{x}^j)\Sigma(x)
\end{equation}
\noindent We note that $\Pi$ and $\Sigma$ are {\it peculiar} velocity dispersions.
Finally, $\zeta_{123}$ is taken to be the hierarchical form (DP; RF)\\
\setlength{\mathindent}{5.5 cm}
\begin{equation}
\zeta_{123} = Q(\xi_{12}\xi_{13} + \xi_{13}\xi_{23} + \xi_{12}\xi_{23})
\end{equation}
\noindent This form for $\zeta_{123}$ satisfies the necessary symmetry under the exchange 
of indices; further, it vanishes when any of the three points is removed to a large 
distance. Using the above ansatz, we obtain \\
\setlength{\mathindent}{5.8 cm}
\begin{equation}
\label{continuity}
{{{\partial \xi}\over{\partial{t}}}} + {1\over{ax^2}}{{\partial\over{\partial{x}}}}
{\Big[} x^2v(1+\xi){\Big]} = 0 
\end{equation}
\noindent and \\
\setlength{\mathindent}{0.0 cm}
\begin{displaymath}
{1\over a}{{\partial\over{\partial{t}}}}{\Big[}av(1+\xi){\Big]} + {1\over {ax^2}}
{{\partial\over{\partial{x}}}}{\Big[}x^2(\Pi+v^2)(1+\xi){\Big]} - {2\over{ax}}\Sigma(1+\xi) 
\end{displaymath}
\setlength{\mathindent}{2.5 cm}
\begin{equation}
\label{euler}
+ {{2G{\rho_b}a}\over {x^2}}\int_0^x d^3z\xi(z) + 2GQ{\rho_b}a\int d^3z{\Big [} \xi(x) 
+ \xi(z){\Big ]}\xi({\bf z}-{\bf x}){{\mbox{cos}\;\theta} \over {z^2}} = 0
\end{equation}
\noindent We next differentiate equation (\ref{continuity}) with respect to $t$ and equation 
(\ref{euler}) with respect to $x$; the resulting equations can be combined to yield\\
\setlength{\mathindent}{0.0 cm}
\begin{displaymath}
-{{{\partial}\over{\partial{t}}}}{\Big [}a^2 {{{\partial (x^2\xi)}\over{\partial{t}}}}{\Big ]} + 
{{{\partial}^2}\over{\partial{x^2}}}{\Big [} x^2(\Pi + v^2)(1 + \xi){\Big ]} - 
{{{\partial}\over{\partial{x}}}}{\Big [} 2\Sigma x(1+\xi){\Big ]} 
\end{displaymath}
\setlength{\mathindent}{5.7 cm}
\begin{equation}
+ {{{\partial}\over{\partial{x}}}}{\Big [} 2G\rho_b a^2\int_0^x d^3z\xi(z){\Big ]} 
 + {{{\partial}\over{\partial{x}}}} {\Big [} 2GQ{\rho_b}a^2x^2M{\Big ]} = 0
\end{equation}
\noindent where $M$ is defined by \\
\setlength{\mathindent}{5.3 cm}
\begin{equation}
 M = \int d^3z{\Big [} \xi(x) + \xi(z){\Big ]}\xi({\bf z}-{\bf x})
{{\mbox{cos}\;\theta} \over {z^2}}
\end{equation}
\noindent Now, the mean 2-point correlation function, $\overline{\xi}(x,a)$, is defined by \\
\setlength{\mathindent}{5.5 cm}
\begin{equation}
\label{barxi}
\overline{\xi}(x,a) = {3 \over {x^3}}\int_0^x dx \xi(x,a) x^2 $$
\end{equation}
\noindent Substituting for $\xi$ in terms of $\overline\xi$, we obtain \\
\setlength{\mathindent}{0.0 cm}
\begin{displaymath}
-{{{\partial}\over{\partial{t}}}}{\Bigg [}a^2 {{{\partial}\over{\partial{t}}}} 
{\Big \{}{{{\partial}\over{\partial{x}}}}{\Big (} {{x^3\overline\xi}\over 3} 
{\Big )}{\Big \}} {\Bigg ]} + {{{\partial}^2}\over{\partial{x^2}}}{\Bigg [} 
( \Pi + v^2) {{{\partial}\over{\partial{x}}}} {\Big [} {{x^3(1+\overline\xi)}
\over 3} {\Big ]} {\Bigg ]} - {{{\partial}\over{\partial{x}}}}{\Bigg [} 
{{2\Sigma} \over x} {{{\partial}\over{\partial{x}}}} {\Big [} {{x^3
(1+\overline\xi)}\over 3} {\Big ]} {\Bigg ]}
\end{displaymath}
\setlength{\mathindent}{6.5 cm}
\begin{equation} 
+ {{{\partial}\over{\partial{x}}}}{\Big [} {{8\pi G}\over 3}\rho_b a^2 x^3\overline\xi {\Big ]} 
+ {{{\partial}\over{\partial{x}}}} {\Big [} 2GQ{\rho_b}a^2x^2M{\Big ]} = 0
\end{equation}
\noindent After integrating over $x$ and carrying out some algebra, this gives \\
\setlength{\mathindent}{0.0 cm}
\begin{equation}
{{{\partial}^2}\overline\xi\over{\partial{A^2}}} + {1 \over 2} {{{\partial{\overline\xi}}
\over{\partial{A}}}} - 3\overline\xi - {\Big (} h^2_{\parallel} + h^2 {\Big )}
{\Big [} 4F + {{{\partial F}\over{\partial{X}}}} {\Big ]} - F {{{\partial}\over{\partial{X}}}} 
{\big [} h^2_{\parallel} + h^2 {\big ]} + 2 h^2_{\perp}F = {{9QMe^{-X}}\over {4\pi}}
\end{equation}
\noindent In the above, we have defined $F$ by \\
\setlength{\mathindent}{5.5 cm}
\begin{equation}
F = x{{{\partial {\overline\xi}}\over{\partial{x}}}} + 3(1 + \overline\xi)
\end{equation}
\setlength{\mathindent}{4.5 cm}
\noindent , substituted $X = \mbox{ln}\hskip 0.03 in x$ and $A = \mbox{ln}
\hskip 0.03 in a$, and defined \\
\setlength{\mathindent}{4.0 cm}
\begin{equation}
h = -{v \over {{\dot a}x}}, \hskip 0.2 in h^2_{\parallel} = {{\Pi} \over {{\dot a}^2x^2}} 
\hskip 0.2 in \mbox{and}  \hskip 0.2 in h^2_{\perp} = {{\Sigma} 
\over {{\dot a}^2x^2}}$$
\end{equation}
\noindent However, ${{{\partial{\overline\xi}}/{\partial{A}}}} = hF $ (NP). Hence, \\
\setlength{\mathindent}{0.0 cm}
\begin{equation}
F{\Big (} {{{\partial h}\over{\partial{A}}}} - h {{{\partial h}\over{\partial{X}}}} {\Big )} 
+ {{hF}\over 2} - h^2F -3\overline\xi - h^2_{\parallel} {\Big (} 4F + 
{{{\partial F}\over{\partial{X}}}} {\Big )} - F {{{\partial h^2_{\parallel}}\over{\partial{X}}}} 
+  2 h^2_{\perp}F = {{9QMe^{-X}}\over {4\pi}}
\end{equation}
\noindent Here, we make the ansatz, $h \equiv h(\overline\xi)$ (\cite{Ham}; NP; \cite{Mo}; 
\cite{PadEng}). This gives \\
\setlength{\mathindent}{0.5 cm}
\begin{equation}
\label{hpartial}
{{{\partial h}\over{\partial{\overline\xi}}}}{\Big [} {{{\partial 
\overline\xi}\over{\partial{A}}}} - h {{{\partial \overline\xi}\over{\partial{X}}}} 
{\Big ]} + {h \over 2} - h^2 - {{3\overline\xi} \over F}
- h^2_{\parallel} {\Big (} 4 + {{{\partial \mbox{ln} \hskip 0.03 in F}\over{\partial{X}}}} {\Big )} - 
{{{\partial h^2_{\parallel}}\over{\partial{X}}}} +  2 h^2_{\perp} = {{9MQe^{-X}}\over {4\pi F}}
\end{equation}
\noindent and, finally, \\
\setlength{\mathindent}{1.0 cm}
\begin{equation}
\label{hevolution}
3h(1 + \overline\xi) {{dh}\over {d\overline\xi}} + {h \over 2} - h^2 - 
{{3\overline\xi} \over F} - h^2_{\parallel} {\Big (} 4 
+ {{{\partial \mbox{ln}\hskip 0.03 in  F}\over{\partial{X}}}} {\Big )} - 
{{{\partial h^2_{\parallel}}\over{\partial{X}}}} 
+  2 h^2_{\perp} = {{9MQe^{-X}}\over {4\pi F}}
\end{equation}
\noindent where we have used (NP) \\
\setlength{\mathindent}{6.0 cm}
\begin{equation}
{{{\partial \overline\xi}\over{\partial{A}}}} - h {{{\partial 
\overline\xi}\over{\partial{X} }}} = 3h( 1 + \overline\xi)
\end{equation}
\noindent Further, the ansatz, $h \equiv h(\overline\xi)$ has 
been used to convert the partial derivative into a total derivative. 
Equation (\ref{hevolution}) governs the evolution of $h$ in terms of 
the mean 2-point correlation function $\overline\xi$. 
\par
\section{The non-linear regime}
\noindent We will consider the small separation, strong clustering limit, $\overline\xi 
\gg 1$. In this regime, the 2-point correlation function was found by DP to exhibit 
a scale-invariant, power-law behaviour, with the assumption of stable clustering. The 
latter is physically well-motivated as it seems reasonable to expect stable, bound 
systems to form under the influence of gravity. Such systems would neither expand 
nor contract and would hence have peculiar velocities equal and opposite to the Hubble 
expansion, {\it i.e.} $v^i_{pec}=-{\dot a}x^i$. The stable clustering ansatz has, however, 
not been deduced from any fundamental considerations and might certainly be considered 
suspect if mergers of structures are important (\cite{Padetal96}). Also, while 
N-body simulations (\cite{Ham}) indicate that $h \rightarrow 1$ for $\xi \gg 1$, 
the results are, at best, inconclusive. The argument can, however, be generalised 
(\cite{Pad97}) as the function $h$, which is the ratio of two velocities, should 
tend to some constant value, if the virialised systems have reached stationarity 
in the statistical sense. We will use this more general stability condition, 
namely $h \rightarrow \mbox{const.}$, to proceed; using this assumption, equation 
(\ref{initcontinuity}) reduces to \\
\setlength{\mathindent}{6.0 cm}
\begin{equation}
a{{\partial\xi}\over{\partial{a}}} - {h\over {x^2}}{\partial\over{\partial{x}}}{\Big[}{x^3} \xi{\Big]}= 0
\end{equation}
\noindent The above equation has the power-law solution, $\xi \sim a^{\beta}x^{-\gamma}$ 
with $\beta = (3-\gamma)h$. Dimensional analysis of equation (\ref{euler}) reveals that $\Pi$ 
and $\Sigma$ scale as $\Pi \sim \Sigma \sim a^{(3-\gamma)h - 1}x^{2-\gamma}$. Now, for a 
flat ($\Omega = 1$) Universe, ${\dot a}^2 \propto a^{-1}$. This implies that \\
\setlength{\mathindent}{5.2 cm}
\begin{eqnarray}
h_{\parallel}^2 = {{\Pi} \over {{\dot a}^2x^2}} &
		\propto  a^{(3-\gamma)h}x^{-\gamma}  &
		= {\Pi_o}\overline\xi
\end{eqnarray}
\noindent and, similarly, \\
\setlength{\mathindent}{7.0 cm}
\begin{equation}
h_{\perp}^2={\Sigma_o}\overline\xi
\end{equation}
\noindent where $\Pi_o$ and $\Sigma_o$ are constants of proportionality. Also, 
(\cite{YG}) we can write $M = M'x\overline\xi^2$, where $M'$ is another constant. 
In the non-linear limit, $\overline\xi \gg 1$, \\
\setlength{\mathindent}{6.5 cm}
\begin{equation}
 F = (3-\gamma)\overline\xi + 3 
\end{equation}
\noindent and \\
\setlength{\mathindent}{5.5 cm}
\begin{equation}
{{{\partial \mbox{ln} \hskip 0.03 in F}\over{\partial{X}}}} = -\gamma
{\Big[} 1 + ({3 \over {3 - \gamma}})({1 \over \overline\xi}) {\Big]}^{-1}$$
\end{equation}
\setlength{\mathindent}{3.5 cm}
\noindent Substituting for $F,h_{\parallel}^2$, $h_{\perp}^2$ and $\overline\xi$ in 
equation (\ref{hevolution}) and using the limit $\overline\xi \gg 1$, we obtain\\
\begin{equation}
\label{ceqn}
3h{\overline\xi}{{dh}\over {d\overline\xi}} - h^2 + {h\over 2} - 
({3 \over {3 -\gamma}})D(\gamma) = {\overline\xi}C(\gamma) + \vartheta({1\over \overline\xi})
\end{equation}
\noindent where $D(\gamma)$ and $C(\gamma)$ are defined by 
\setlength{\mathindent}{5.0 cm}
\begin{equation}
D(\gamma) = 1 + \gamma \Pi_o - {{9QM'} \over {4\pi(3-\gamma)}}
\end{equation}
\noindent and \\
\setlength{\mathindent}{4.5 cm}
\begin{equation}
C(\gamma) = 2\Pi_o(2-\gamma) - 2\Sigma_o + {{9QM'} \over {4\pi(3-\gamma)}}
\end{equation}
\noindent Equation (\ref{ceqn}) is exact (within the exact power law solutions for 
$\overline\xi$, $h_{\parallel}^2$, $h_{\perp}^2$ and $M$) upto order constant, 
in the limit of large $\overline\xi$, with terms of order $\vartheta(1/\overline\xi)$ 
neglected. \\
\noindent Clearly, $C(\gamma)$ must be exactly zero, as otherwise, $h \propto 
\sqrt{\overline\xi}$ for $\overline\xi \gg 1$, which violates the stable 
clustering hypothesis. Note that $C(\gamma) \approx 0$ is not sufficient 
as, if this were the case, the term in $C(\gamma)$ would cause $h$ to grow with 
$\overline\xi$, for sufficiently large $\overline\xi$. Thus \\
\setlength{\mathindent}{5.0 cm}
\begin{equation}
\label{czero}
2\Sigma_o - 2\Pi_o(2-\gamma) = {{9QM'} \over {4\pi(3-\gamma)}}
\end{equation}
\noindent This equation is equivalent to equation (48) of RF, albeit in slightly 
different form. Equation (\ref{ceqn}) thus reduces to \\
\setlength{\mathindent}{5.5 cm}
\begin{equation}
\label{hfinal}
3h{\overline\xi}{{dh}\over {d\overline\xi}} - h^2 + {h\over 2} = A 
\end{equation}
\noindent where $A = [3/(3 - \gamma)]D(\gamma)$.\\
\noindent Now, the ansatz of stable clustering implies that 
$\overline\xi dh/d\overline\xi \rightarrow 0$ as $\overline\xi 
\rightarrow \infty$. Equation (\ref{hfinal}) then gives \\
\setlength{\mathindent}{6.0 cm}
\begin{equation}
h^2 - {h\over 2} + A = 0
\end{equation}
\noindent {\it i.e.} \\
\setlength{\mathindent}{5.5 cm}
\begin{equation}
\label{hval}
h = {1\over 4}\left[ 1 \pm \sqrt{1 - 16 A} \right]
\end{equation}
\noindent Since $h$ is a real quantity, the above equation immediately 
yields \\
\setlength{\mathindent}{7.0 cm}
\begin{equation}
 A \leq {1\over 16}
\end{equation}
\noindent This gives \\
\setlength{\mathindent}{5.0 cm}
\begin{equation}
1 + \gamma \Pi_o - {{9QM'} \over {4\pi(3-\gamma)}} \leq \left( 
{{3 - \gamma}\over 48} \right)
\end{equation}
\noindent Replacing for ${{9QM'}/ {4\pi(3-\gamma)}}$ from equation (\ref{czero}), 
we obtain \\
\setlength{\mathindent}{5.0 cm}
\begin{equation}
\label{ineq}
2 \Sigma_o \geq (4 -\gamma)\Pi_o + 1 - \left( {{3 - \gamma}\over 48} \right)
\end{equation}
\noindent RF have pointed out that various integrals in the BBGKY hierarchy do not 
converge unless $0 < \gamma < 2$. This range of $\gamma$ values implies that 
$1 -  {{(3 - \gamma)}/48} > 0$ and $(4 - \gamma) > 2$. Thus, the inequality 
(\ref{ineq}) gives \\
\setlength{\mathindent}{7.0 cm}
\begin{equation}
\Sigma_o > \Pi_o
\end{equation}
\noindent Thus, the DP solution requires that tangential dispersions exceed radial 
dispersions in the non-linear regime. This is understandable on physical grounds, 
as tangential dispersions cause deviations from radial infall; stable structures would 
hence only be expected to form once these dispersions become comparable to or larger 
than the radial ones. \\
\noindent The general solution to equation (\ref{hfinal}) has the following 
three different forms depending on whether $(1 - 16A)$ is positive, negative or 
zero. The constant of integration is denoted by $B$ in each case. \\

\noindent 1. For $(1 - 16A) < 0 $, the solution is \\
\setlength{\mathindent}{1.0 cm}
\begin{equation}
\label{nosoln}
{\rm ln}\left( {B\overline\xi^{1/6}} \right) = {\rm ln}\left[\left(  2h^2 - h
+ 2A
\right)^{1/4} \right] + {1 \over 2} \sqrt{{{ 3 - \gamma} \over { \gamma + 45}}}
{\rm Tan}^{-1}\left[ \left( 4h - 1 \right) \sqrt{{{ 3 - \gamma }\over {\gamma
+ 45}}}
\right]
\end{equation}
\noindent 2. Next, for $A = 1/16$, the solution has the form \\
\setlength{\mathindent}{5.0 cm}
\begin{equation}
\label{a16}
B{\overline\xi}^{2/3} = \left[ 2h^2 - h + 2A \right] {\rm exp} \left[ -{2\over
 {4h-1}}\right]
\end{equation}
\noindent 3. Finally, for $(1 - 16A) > 0$, the solution is \\
\setlength{\mathindent}{5.0 cm}
\begin{equation}
\label{usesoln}
B{\overline\xi^{2/3}} = \left[ 2h^2 - h + 2A \right] \left[ {{p - 4h + 1}\over
{p + 4h - 1}} \right]^{1/p}
\end{equation}
\noindent where we have defined $ p = \sqrt{1 - 16A}$. As mentioned earlier, 
$(1 - 16A)$ cannot be negative since $h$ is a real quantity; equation (\ref{nosoln}) 
can hence be immediately ruled out as a possible solution. Further, although 
$\overline\xi dh/d\overline\xi\rightarrow 0$, $h = 0.25(1 \pm |{p}|)$ is certainly 
a solution of equation (\ref{hfinal}), it is not clear if this can be actually 
reached from the general solution embodied in equations (\ref{a16}) and 
(\ref{usesoln}). We hence perturb these solutions by writing $h = h_o + 
\epsilon$, where $\epsilon$ is the perturbation parameter and $h_o$ satisfies 
the equation \\
\setlength{\mathindent}{6.5 cm}
\begin{equation}
\label{hoeqn}
h_o^2 - {h_o\over 2} =  {3 \over {\gamma - 3}}
\end{equation}
\noindent {\it i.e.}
\begin{equation}
\label{hoval}
h_o = 0.25\left( 1 \pm |p|\right)
\end{equation}
\noindent We then attempt to impose the condition $\epsilon \rightarrow 0$
as $\overline\xi \rightarrow \infty$; if this is possible, it clearly indicates 
that $h \rightarrow h_o$ as $\overline\xi \rightarrow \infty$, {\it i.e.} a
 solution exists which satisfies the stable clustering hypothesis. We initially 
consider the case $A = 1/16$ ($p = 0$) and rewrite equation (\ref{a16}) as \\
\setlength{\mathindent}{3.0 cm}
\begin{equation}
B{\overline\xi}^{2/3} = \left[ 2h_o^2- h_o + 2A +4h_o\epsilon + \epsilon^2 - 
\epsilon \right] {\rm exp} \left[ {2\over {1- 4h-4\epsilon}} \right]
\end{equation}
\noindent Since $h_o = 0.25$ for $p = 0$, this gives \\
\setlength{\mathindent}{6.5 cm}
\begin{equation}
B{\overline\xi}^{2/3} = \left[ \epsilon^2 \right] {\rm exp} \left[ -{1\over
{2\epsilon}} \right]
\end{equation}
\noindent One can impose the condition $\epsilon \rightarrow 0$ as $\overline\xi 
\rightarrow \infty$, in the above equation, only if $\epsilon$ is negative. 
$ A = 1/16$ is thus an allowed solution and $h \rightarrow 0.25$ from below, as 
$\overline\xi \rightarrow \infty$. \\
\noindent Finally, we consider the case $(1 - 16A) > 0$, {\it i.e.} $A < 1/16$. 
Equation (\ref{usesoln}) can be rewritten as \\
\setlength{\mathindent}{2.5 cm}
\begin{equation}
B{\overline\xi^{2/3}} = \left[ 2h_o^2 - h_o + 2A + 2\epsilon^2 + 4h_o \left( 
\epsilon - 1 \right) \right] \left[ {{p - 4h_o - 4\epsilon  + 1}\over {p + 4h_o + 
4\epsilon - 1}} \right]^{1/p}
\end{equation}
\noindent Using equation (\ref{hoeqn}) in the above and retaining terms upto
first order in $\epsilon$, we obtain \\
\setlength{\mathindent}{4.5 cm}
\begin{equation}
\label{pertsoln}
B{\overline\xi^{2/3}} = \left[ 4h_o \left( \epsilon- 1 \right) \right]
\left[ {{p - 4h_o - 4\epsilon  + 1}\over {p + 4h_o +4\epsilon - 1}} \right]^{1/p}
\end{equation}
\noindent The two possible values of $h_o$, given by equation (\ref{hoval}), are
$h_o = 0.25\left( 1 \pm |p|\right)$. For $h_o = 0.25\left( 1 - |p|\right)$, we choose 
$p > 0$ in equation (\ref{pertsoln}). This gives \\
\setlength{\mathindent}{5.0 cm}
\begin{eqnarray}
B{\overline\xi^{2/3}}&=& -|p| \epsilon \left[ {{2|p| - 4\epsilon} \over 
{ 4\epsilon}} \right]^{1/|p|}\\
\label{goodsoln}
&=&-|p|\left[ {{|p| - 2\epsilon}\over 2} \right]^{1/|p|} \epsilon^{1 - 1/|p|}
\end{eqnarray}
\noindent Equation (\ref{goodsoln}) shows that one can satisfy the the condition
$\epsilon \rightarrow 0$ as $\overline\xi \rightarrow \infty$ if $1 < |p|^{-1}$
{\it i.e.} if \\
\setlength{\mathindent}{7.0 cm}
\begin{equation}
\sqrt{1 - 16A} < 1
\end{equation}
\noindent or, in other words, $A > 0$. (An equivalent result can be obtained 
for the solution $h_o = 0.25\left( 1 + |p|\right)$ by choosing $p < 0$ in 
equation (\ref{pertsoln})).\\
\noindent Since $A \geq 0$, equation (\ref{hval}) then implies that \\
\begin{equation}
h(h - {1\over 2}) \leq 0
\end{equation}
\noindent {\it i.e.} $ 0 \leq h \leq 1/2$. Thus, the asymptotic values of $h$, allowed by 
the DP solution and consistent with the stable clustering hypothesis, lie in the 
range $0 < h < 1/2$. In the standard stable clustering scenario, $h \rightarrow 1$ 
for $\overline\xi \gg 1$. The above analysis clearly rules out this value of $h$ as 
$\overline\xi \rightarrow \infty$.
\noindent Next, the constraint $A \geq 0$ gives \\
\setlength{\mathindent}{6.0 cm}
\begin{equation}
1 + \gamma\Pi_o - {{9QM'} \over {4\pi(3-\gamma)}} \geq 0 
\end{equation}
\noindent Using equation (\ref{czero}), this gives \\
\begin{equation}
1 + \left( 4 - \gamma \right) \Pi_o - 2\Sigma_o \geq 0
\end{equation}
\noindent {\it i.e.} \\
\setlength{\mathindent}{6.5 cm}
\begin{equation}
\label{sigless}
\Sigma_o \leq {{\left( 4 - \gamma \right)}\over 2} \Pi_o + {1\over 2}
\end{equation}
\noindent We have already shown that $\Sigma_o > \left[{{\left( 4 - \gamma 
\right)}/2}\right] 
\Pi_o + {1/ 2} - (3 - \gamma)/96$. The above equation (\ref{sigless}) indicates 
that only a very narrow range of values is permitted for $\Sigma_o$ (in terms of 
$\Pi_o$), in the non-linear regime. Thus, the DP solution imposes strong constraints 
on the relative values of radial and tangential dispersions; it is not obvious that 
the dynamics of the system will actually cause these constraints to be satisfied. \\
\noindent Finally, the preceding results do not appear to be influenced by the ansatz 
$h \equiv h(\overline\xi)$ (although this ansatz is well-motivated, since it appears 
to be validated in N-body simulations; see, for example, \cite{Ham}) as one could 
have instead carried out the analysis using the scaled variable $s \equiv xt^{-\alpha}$ 
(with $\gamma = 2/(\alpha + 2/3)$). In the non-linear regime, self-similar evolution 
implies that $h \equiv h(s)$ and $\bar\xi \equiv \bar\xi(s)$; one can clearly write 
$h \equiv h(\bar\xi)$ in this regime.\\
\section{Conclusions}
\noindent In the present work, the BBGKY hierarchy of equations has been revisited 
and an equation derived for the evolution of the dimensionless function $h = -({v 
/ {\dot{a}x}})$. The assumptions used (following DP) are a hierarchical model for the 
3-point correlation function and the ansatz that $h$ is a function of the mean 
2-point correlation function, $\overline\xi$, alone, {\it i.e.} $h \equiv h(\overline\xi)$. 
No assumption is made regarding the vanishing of velocity skewness. In fact, the 
analysis only uses the zeroth and first moments of the second BBGKY equation; 
it is thus applicable to any form of the skewness (or the higher velocity moments)
which yields the DP solution for $\overline\xi$ and the parallel and perpendicular 
velocity dispersions. The DP similarity solution is then substituted in this
equation for the $h$ function, in the non-linear regime, and the generalised stable 
clustering hypothesis ($h \rightarrow \mbox{const.}$) used to obtain an expression 
for the asymptotic value of $h$, in terms 
of $\gamma$, the power law index of clustering and the tangential and radial 
velocity dispersions. The DP solution is found to require that tangential 
dispersions are larger than radial ones, in the strong clustering regime; this can be 
understood on physical grounds. Finally, stability analysis of the solution 
demonstrates that the allowed asymptotic values of $h$, consistent with the stable 
clustering hypothesis, lie in the range $0 \leq h \leq 1/2$. Thus, if the DP 
scale-invariant solution (and the hierarchical model for the 3-pt function) 
is correct, the standard stable clustering picture ($h \rightarrow 1$ as 
$\overline\xi \rightarrow \infty$) is not allowed in the non-linear regime 
of structure formation.\\ 
\par
\acknowledgments{It is a pleasure to thank T. Padmanabhan for initially 
suggesting this problem to me. I also thank him and Kandu Subramanian 
for exceedingly useful discussions during the course of the present 
work, as well as for a critical reading of previous drafts of this paper. 
Finally, I thank an anonymous referee for his/her comments on and criticism of 
an earlier version.}

\end{document}